# Path-Integral Renormalization Group Treatments for Many-Electron Systems with Long-Range Repulsive Interactions

Masashi Kojo[*] and Kikuji Hirose

Graduate School of Engineering, Osaka University, 2-1 Yamadaoka, Suita, Osaka, 565-0871, Japan

Phone/Fax: +81-6-6879-7290, E-mail: kojo@cp.prec.eng.osaka-u.ac.jp

A practical algorithm for many-electron systems based on the path-integral renormalization group (PIRG) method is proposed in the real-space finite-difference (RSFD) approach. The PIRG method, developed for investigating strongly correlated electron systems, has been successfully applied to some models such as Hubbard models. However, to apply this method to more realistic systems of electrons with long-range Coulomb interactions within the RSFD formalism, the one-body Green's function, which requires large computational resources, is to be replaced with an alternative. For the same reason, an efficient algorithm for computing the Fock matrix is needed. The newly proposed algorithm is free of the one-body Green's function and enables us to compute the Fock matrix efficiently. Our result shows a significant reduction in CPU time and the possibility of using the present algorithm as a practical numerical tool.

## INTRODUCTION

Up to now, many numerical algorithms for strongly correlated electron systems have been proposed and applied to various systems, e.g., quantum Monte Carlo method [1], density matrix renormalization group (DMRG) method [2], configuration interaction method [3], and coupled cluster method [4]. However, even now, the nature of the ground state still remains a challenge because of the immaturity of numerical tools.

The path-integral renormalization group (PIRG) method [5, 6] was proposed to obtain the many-body ground state. Unlike the quantum Monte Carlo method, the PIRG method does not suffer from the sign problem [7]. Furthermore, unlike the DMRG method, the PIRG method does not limit the dimensionality of systems because numerical renormalization is carried out in an imaginary time space. In the PIRG method, the ground-state wave function is expressed by a linear combination of basis states, e.g., Slater determinants, in a truncated Hilbert space. While retaining the size of the truncated Hilbert space, the optimized basis states and the ground state are projected out numerically.

To make the PIRG method applicable to more realistic systems, we extend the PIRG method with the real-space finite-difference (RSFD) approach in which every physical quantity is defined only on grid points in the discretized space [8-11]. In this endeavor, The process of "choosing more preferable basis states" becomes the main drawback with respect to the computational cost. In particular, because more basis states tend to be required in the RSFD scheme, one-body Green's functions and the Fock matrix dominate the computational cost and prevent us from applying this method to realistic systems.

The aim of this work is to introduce a new algorithm within the framework of the RSFD approach to overcome the above-mentioned problems and to show its applicability to a realistic quantum system.

## METHODOLOGY AND NUMERICAL APPLICATION

### Wave Function Representation

Let us expand the many-body wave function

$|\Psi\rangle$ in terms of *non-orthogonal* Slater determinants $\{|\Phi_l\rangle\}$ [5, 6], i.e.,

$$|\Psi\rangle = \sum_{l=1}^{L} w_l |\Phi_l\rangle,$$

$$|\Phi_l\rangle = \prod_{m=1}^{M}\left(\sum_{j=1}^{N_{grid}}[\Phi_l]_{jm}\hat{c}_j^\dagger\right)|0\rangle,$$

$$[\Phi_l] = (\vec{\phi}_{l1}, \cdots, \vec{\phi}_{lM}),$$

Where $\{w_l\}$ are expansion coefficients, $\hat{c}_j^\dagger$ is the creation operator of electrons at the j-th grid point, $\{\vec{\phi}_{lm}\}$ are column vectors of the matrix representing Slater determinants, and $L$, $M$ and $N_{grid}$ denote the numbers of basis states, electrons and grid points in the discretized space, respectively. Note that $N_{grid}$ is taken to be extremely larger than $M$ in the RSFD formalism. Throughout this paper we ignore spin indices for simplicity.

**Imaginary-Time Evolution of Systems**

The process of "choosing more preferable basis states" involves the operation of the imaginary-time propagator to Slater determinants and choosing the best basis set that gives a lower expectation value of energy than the other sets. Operating the imaginary-time propagator leads the ground state to be projected out. By taking sufficiently small $\delta\tau$ and imaginary time $\tau$, this projection process is written as

$$|\Psi_{\tau+d\tau}\rangle = \exp(-\delta\tau\hat{H})|\Psi_\tau\rangle.$$

The Hamiltonian which we tackle is

$$\hat{H} = \hat{H}_K + \hat{H}_{ext} + \hat{H}_{ee},$$

$$\hat{H}_K = \sum_{j=1}^{N_{grid}} \hat{c}_j^\dagger \left(-\frac{1}{2}\Delta\right) \hat{c}_j,$$

$$\hat{H}_{ext} = -\sum_{k=1}^{N_{nuc}}\sum_{j=1}^{N_{grid}} \hat{c}_j^\dagger Z_k v_{j,nuc(k)} \hat{c}_j,$$

$$\hat{H}_{ee} = \frac{1}{2}\sum_{j,k=1}^{N_{grid}} \hat{c}_j^\dagger \hat{c}_k^\dagger v_{jk} \hat{c}_k \hat{c}_j,$$

where

$$v_{jk} = \frac{1}{|\vec{r}_j - \vec{r}_k|},$$

and $N_{nuc}$ denote the number of nuclei, and $Z_k$ and $nuc(k)$ are the atomic number and location of the k-th nucleus, respectively. Here and hereafter, we adopt the central finite-difference formula [8, 9] for the Laplacian, i.e., $\Delta\hat{c}_j$ means $(\hat{c}_{j+1} - 2\hat{c}_j + \hat{c}_{j-1})/\delta r^2$ in each direction, where $\delta r$ is the grid spacing.

Because the propagator contains two-body interactions, operating the propagator to Slater determinants generates enormous numbers of Slater determinants. To avoid this problem, two-body operators in the propagator are decomposed into one-body operators by Suzuki-Trotter decomposition,

$$\exp(-\delta\tau\hat{H}) = \exp(-\delta\tau\hat{H}_K/2)$$
$$\times \exp(-\delta\tau\hat{H}_{ext})\exp(-\delta\tau\hat{H}_{ee})$$
$$\times \exp(-\delta\tau\hat{H}_K/2) + O(\delta\tau^3),$$

and an auxiliary field technique. See Appendix A for details of the latter technique, which is summarized as

$$\exp(-\delta\tau\hat{H}_{ee})$$
$$= Z_A^{-1} \int \prod_{j=1}^{N_{grid}} dA_j \exp\left\{-\delta\tau\left[S_A + \hat{H}'_{ee}(A_j)\right]\right\},$$

$$\hat{H}'_{ee}(A_j) = \sum_{j=1}^{N_{grid}} \hat{c}_j^\dagger (-iA_j - v_{jj}/2)\hat{c}_j,$$

$$S_A = \delta r^3 \sum_{j=1}^{N_{grid}} \frac{1}{8\pi}\left\{(\nabla A_j)^2 + (\mu A_j)^2\right\},$$

$$Z_A = \int \prod_{j=1}^{N_{grid}} dA_j \exp(-\delta\tau S_A).$$

Here, $A_j$ represents an auxiliary field variable and $\mu$ denotes the parameter concerning the screened Coulomb potential [see Eq. (3)].

To treat the kinetic part in the imaginary-time propagator, the following approximation is frequently used:

$$\exp(-\delta\tau\hat{H}_K/2) \approx 1 - \frac{\delta\tau\hat{H}_K}{2} + O(\delta\tau^2). \quad (1)$$

Although the PIRG method is stable to accumulated numerical errors that originate in the above equation, the required number of time steps to achieve a convergence may be affected. Hence we employ the following exact formulation instead of Eq. (1):

$$\langle x_l | \exp(-\delta\tau \hat{H}_K/2) | x_m \rangle$$
$$= N_{grid}^{-1} \sum_{n=1}^{N_{grid}} \exp[iG_n(l-m)\delta r - \delta\tau\varepsilon_n/2], \quad (2)$$

where
$$G_n = \frac{2\pi n}{N_{grid}\delta r} \quad \text{and} \quad \varepsilon_n = \frac{2}{\delta r^2}\sin^2(G_n\delta r/2).$$

For simplicity, Eq. (2) is based on the assumption of one-dimensional periodic systems. The proof is given in Appendix B. Fig. 1 illustrates the kinetic propagator calculated using Eq. (2) in which $\delta r = \delta\tau = 0.1$ a.u. and $N_{grid} = 61$. The locality of the kinetic propagator enables us to treat it efficiently, as in the case of Eq. (1).

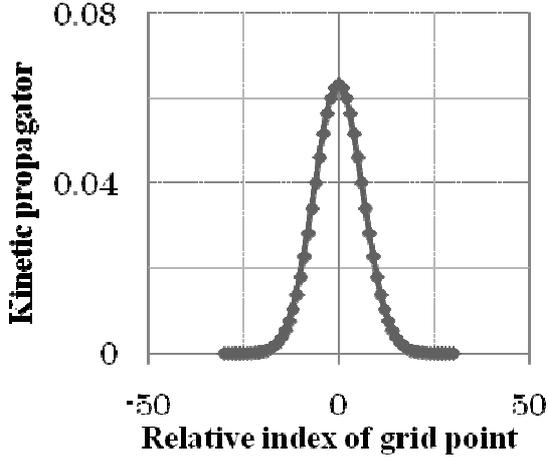

Fig. 1. Exact kinetic propagator

**Discretized Screened Coulomb Potential**

Since potentials are defined on grid points in the RSFD approach, a discretized Coulomb potential is used instead of the continuous one. In order to avoid numerical difficulties, the Helmholtz equation is employed instead of the Poisson equation. In the continuous case, the Helmholtz equation and its solution, i.e., the screened Coulomb potential (Yukawa potential), are expressed by
$$(\Delta - \mu^2)v_{jk} = -4\pi\delta(\vec{r}_j - \vec{r}_k),$$
$$v_{jk} = \frac{\exp(-\mu|\vec{r}_j - \vec{r}_k|)}{|\vec{r}_j - \vec{r}_k|}. \quad (3)$$

Here, $\mu$ is a positive number, the inverse of which is called the screening distance. In the discretized case, by means of the discrete Fourier transformation, Eq. (3) is changed into the following form under the periodic boundary condition:
$$v_{jk} = \sum_{n=1}^{N_{grid}} \lambda_n p_{nj}^* p_{nk}, \quad (4)$$

where
$$p_{nk} = \exp(-i\vec{G}_n \cdot \vec{r}_k),$$
$$\lambda_n = \frac{4\pi}{\Omega}\left(\sum_{d=1}^{3} 2\varepsilon_{n_d} + \mu^2\right)^{-1},$$
$$n = (n_1, n_2, n_3),$$
$$\varepsilon_{n_d} = \frac{2}{\delta r^2}\sin^2(G_{n_d}\delta r/2),$$
$$G_{n_d} = \frac{2\pi n_d}{\Omega_d}, \quad \Omega = \prod_{d=1}^{3}\Omega_d$$

with $\Omega_d$ being the supercell size along the d-coordinate axis.

**Alternative to One-Body Green's Functions**

In the original PIRG method [5, 6], the one-body Green's functions $\langle \hat{c}_j^\dagger \hat{c}_k \rangle$ are used to compute physical expectation values. Because the one-body Green's function requires $O(N_{grid}^2)$ computations and memory spaces with respect to $N_{grid}$, it should be altered in the RSFD formalism. For the same reason, an efficient method of computing the Coulomb interaction parts is needed.

As shown in Appendix C, in our Green's-function-free scheme, the kinetic part in the energy expectation value becomes the following linear combination of the inner product of Slater determinants:
$$\langle \Phi | \hat{H}_K | \Phi \rangle = \sum_{m=1}^{M} \langle \Phi | \Phi^m \rangle.$$

Here,
$$|\Phi^m\rangle = \prod_{k=1}^{M}\left(\sum_{j=1}^{N_{grid}}[\Phi^m]_{jk}\hat{c}_j^\dagger\right)|0\rangle,$$
$$[\Phi^m] = (\vec{\phi}_1, \cdots, [H_K]\vec{\phi}_m, \cdots, \vec{\phi}_M),$$
$$[H_K]_{ij} = \langle \vec{r}_i | \hat{H}_K | \vec{r}_j \rangle.$$

Note that $\langle \Phi_A | \Phi_B \rangle = \det([\Phi_A]^\dagger[\Phi_B])$ [6]. Similarly, the electron-electron interaction part and the external one are calculated as follows:

$$2\langle\Phi|\hat{H}_{ee}|\Phi\rangle$$
$$=\sum_{n=1}^{N_{grid}}\lambda_n\sum_{m,m'=1}^{M}\langle\Phi_n^m|\Phi_n^{m'}\rangle-M\sum_{n=1}^{N_{grid}}\lambda_n\langle\Phi|\Phi\rangle,$$
$$\langle\Phi|\hat{H}_{ext}|\Phi\rangle=\sum_{n=1}^{N_{grid}}\lambda_n\sum_{m=1}^{M}\langle\Phi|\Phi_n^m\rangle\sum_{k=1}^{N_{nuc}}p_{n,nuc(k)}^*,$$

where

$$|\Phi_n^m\rangle=\prod_{k=1}^{M}\left(\sum_{j=1}^{N_{grid}}[\Phi_n^m]_{jk}\hat{c}_j^\dagger\right)|0\rangle,$$
$$[\Phi_n^m]=(\vec{\phi}_1,\cdots,\vec{p}_n\cdot\vec{\phi}_m,\cdots,\vec{\phi}_M),$$
$$(\vec{p}_n)_k=p_{nk}.$$

The fact that the difference among Slater determinants in this form is just only one column vector and the Fast Fourier Transform technique enable us to compute the energy expectation values with $O(N_{grid}\ln N_{grid})$.

Fig. 2 shows the comparison of the CPU time required to compute the energy expectation value with $M=2$ on the Intel(R) Pentium(R) D CPU 3.20GHz system. A significant reduction can be seen.

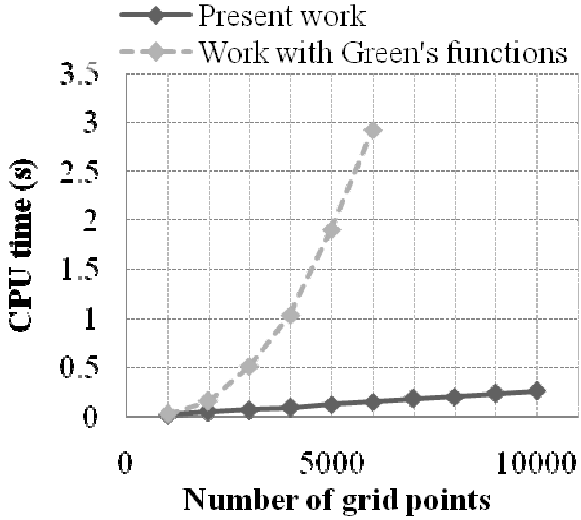

Fig. 2. Comparison of CPU time versus the number of grid points $N_{grid}$

**Numerical Application**

In Fig. 3, we show a preliminary result of the charge distribution of the hydrogen molecule under the three-dimensional periodic boundary condition within the box normalization. At most, 32 basis states and 71 grid points along each direction are taken. The atomic distance, $\delta r$ and $\mu$ are set to 1.3 Å, 0.1 a.u. and 0.1 a.u., respectively. The parallel spin configuration is taken.

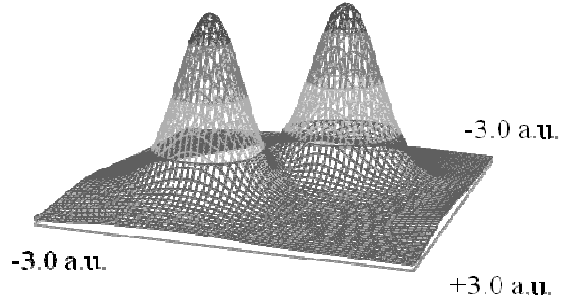

Fig. 3. Charge distribution of hydrogen molecule

**CONCLUSION**

A new algorithm for appling the PIRG method to realistic systems with long-range repulsive interactions is proposed in the RSFD approach. By utilizing this algorithm, the time-consuming one-body Green's functions within the RSFD approach are removed from the PIRG scheme and the Fock matrix can be efficiently computed. Our result shows a significant reduction in CPU time and suggests the possibility of using this algorithm as a practical tool to treat undiscovered systems which cannot be dealt with existing methods.

**ACKNOWLEDGEMENTS**

The authors wish to acknowledge support through a Grant-in-Aid from the 21st Century Center of Excellence (COE) Program "Center for Atomistic Fabrication Technology" and Scientific Research in Priority Areas "Development of New Quantum Simulators and Quantum Design" (Grant No. 17064012) from the Ministry of Education, Culture, Sports, Science and Technology.

**Appendix A: Auxiliary Field Technique**

Consider the transformation of an auxiliary field variable as

$$A_j=A_j'+i\sum_{k=1}^{N_{grid}}v_{jk}\hat{c}_k^\dagger\hat{c}_k.$$

Using this transformation, we obtain the following equations:

$$\frac{1}{8\pi}\delta r^3 \sum_{j=1}^{N_{grid}} \left\{(\nabla A_j)^2 + (\mu A_j)^2\right\}$$

$$= -\frac{1}{8\pi}\delta r^3 \sum_{j=1}^{N_{grid}} A_j(\Delta - \mu^2)A_j$$

$$= -\frac{1}{8\pi}\delta r^3 \sum_{j=1}^{N_{grid}} A'_j(\Delta - \mu^2)A'_j$$

$$+ i\sum_{j=1}^{N_{grid}} A'_j \hat{c}_j^\dagger \hat{c}_j - \frac{1}{2}\sum_{j,k=1}^{N_{grid}} v_{jk} \hat{c}_j^\dagger \hat{c}_j \hat{c}_k^\dagger \hat{c}_k,$$

and

$$-i\sum_{j=1}^{N_{grid}} A_j \hat{c}_j^\dagger \hat{c}_j = -i\sum_{j=1}^{N_{grid}} A'_j \hat{c}_j^\dagger \hat{c}_j + \sum_{j,k=1}^{N_{grid}} v_{jk} \hat{c}_j^\dagger \hat{c}_j \hat{c}_k^\dagger \hat{c}_k.$$

The sum of these two equations leads to the formula as

$$\exp(-\delta\tau \hat{H}_{ee})$$
$$= Z_A^{-1} \int \prod_{j=1}^{N_{grid}} dA_j \exp\left\{-\delta\tau[S_A + \hat{H}'_{ee}(A_j)]\right\}.$$

**Appendix B: Kinetic Propagator**

For simplicity, we assume a discretized one-dimensional periodic system. In such a system, the Schrödinger equation for free electrons and its solutions are given by

$$-\frac{1}{2}\frac{d^2}{dx^2}\phi_{n,j} = \varepsilon_n \phi_{n,j},$$

$$\varepsilon_n = \frac{2}{\delta r^2}\sin^2(G_n \delta r/2),$$

$$\phi_{n,j} = \frac{1}{\sqrt{N_{grid}}}\exp(iG_n j\delta r),$$

$$G_n = \frac{2\pi n}{N_{grid}\delta r}.$$

Then the kinetic propagator becomes

$$\langle x_l | \exp(-\delta\tau \hat{H}_K/2) | x_m \rangle$$

$$= \langle x_l | \exp(-\delta\tau \hat{H}_K/2) \sum_{n=1}^{N_{grid}} |\phi_n\rangle\langle\phi_n | x_m \rangle$$

$$= \sum_{n=1}^{N_{grid}} \exp(-\delta\tau \varepsilon_n/2) \langle x_l | \phi_n \rangle \langle \phi_n | x_m \rangle$$

$$= N_{grid}^{-1} \sum_{n=1}^{N_{grid}} \exp[iG_n(l-m)\delta r - \delta\tau \varepsilon_n/2].$$

**Appendix C: Green's-Function-Free Scheme**

Using the relation

$$\hat{c}_j^\dagger \hat{c}_k |\Phi\rangle = \hat{c}_j^\dagger \hat{c}_k \prod_{p=1}^{M} \hat{\Theta}_p |0\rangle$$

$$= \sum_{q=1}^{M}(-1)^{q+1}[\Phi]_{kq} \hat{c}_j^\dagger \prod_{\substack{p=1\\p\neq q}}^{M} \hat{\Theta}_p |0\rangle$$

with

$$\hat{\Theta}_p = \sum_{n=1}^{N_{grid}} [\Phi]_{np} \hat{c}_n^\dagger,$$

the kinetic term in the energy expectation value is written as

$$\langle \Phi | \hat{H}_K | \Phi \rangle = \frac{-1}{2}\sum_{j=1}^{N_{grid}} \langle \Phi | \hat{c}_j^\dagger \Delta \hat{c}_j | \Phi \rangle$$

$$= \langle \Phi | \left(\frac{-1}{2\delta r^2} \sum_{j=1}^{N_{grid}} [\Phi]_{j1}^\Delta \hat{c}_j^\dagger\right) \prod_{p=2}^{M} \hat{\Theta}_p |0\rangle$$

$$+ \cdots + \langle \Phi | \prod_{p=1}^{M-1} \hat{\Theta}_p \left(\frac{-1}{2\delta r^2} \sum_{j=1}^{N_{grid}} [\Phi]_{jM}^\Delta \hat{c}_j^\dagger\right) |0\rangle$$

$$= \sum_{m=1}^{M} \langle \Phi | \Phi^m \rangle,$$

where

$$[\Phi]_{jk}^\Delta = [\Phi]_{j+1,k} - 2[\Phi]_{j,k} + [\Phi]_{j-1,k},$$

$$|\Phi^m\rangle = \prod_{k=1}^{M}\left(\sum_{j=1}^{N_{grid}} [\Phi^m]_{jk} \hat{c}_j^\dagger\right)|0\rangle,$$

$$[\Phi^m] = (\vec{\phi}_1, \cdots, [H_K]\vec{\phi}_m, \cdots, \vec{\phi}_M),$$

$$[H_K]_{ij} = \langle \vec{r}_i | \hat{H}_K | \vec{r}_j \rangle.$$

In a similar way, the electron-electron interaction term and the external one become

$$2\langle \Phi | \hat{H}_{ee} | \Phi \rangle = \langle \Phi | \sum_{j,k=1}^{N_{grid}} \hat{c}_j^\dagger \hat{c}_k^\dagger v_{jk} \hat{c}_k \hat{c}_j | \Phi \rangle$$

$$= \sum_{n=1}^{N_{grid}} \lambda_n \langle \Phi | \sum_{j=1}^{N_{grid}} \hat{n}_j p_{nj}^* \sum_{k=1}^{N_{grid}} p_{nk} \hat{n}_k | \Phi \rangle$$

$$- \sum_{n=1}^{N_{grid}} \lambda_n \langle \Phi | \sum_{j=1}^{N_{grid}} \hat{n}_j | \Phi \rangle$$

$$= \sum_{n=1}^{N_{grid}} \lambda_n \sum_{m,m'=1}^{M} \langle \Phi_n^m | \Phi_n^{m'} \rangle - M \sum_{n=1}^{N_{grid}} \lambda_n \langle \Phi | \Phi \rangle,$$

$$\langle \Phi | \hat{H}_{ext} | \Phi \rangle = \langle \Phi | \sum_{k=1}^{N_{nuc}} \sum_{j=1}^{N_{grid}} \hat{c}_j^\dagger v_{j,nuc(k)} \hat{c}_j | \Phi \rangle$$

$$= \sum_{n=1}^{N_{grid}} \lambda_n \langle \Phi | \sum_{j=1}^{N_{grid}} \hat{n}_j p_{nj} | \Phi \rangle \sum_{k=1}^{N_{nuc}} p_{n,nuc(k)}^*$$

$$= \sum_{n=1}^{N_{grid}} \lambda_n \sum_{m=1}^{M} \langle \Phi | \Phi_n^m \rangle \sum_{k=1}^{N_{nuc}} p_{n,nuc(k)}^*,$$

where

$$\left| \Phi_n^m \right\rangle = \prod_{k=1}^{M} \left( \sum_{j=1}^{N_{grid}} [\Phi_n^m]_{jk} \hat{c}_j^\dagger \right) |0\rangle,$$

$$[\Phi_n^m] = (\vec{\phi}_1, \cdots, \vec{p}_n \cdot \vec{\phi}_m, \cdots, \vec{\phi}_M),$$

$$(\vec{p}_n)_k = p_{nk}.$$